\documentclass{nature2}
\usepackage{graphicx}
\usepackage[usenames]{color}
\usepackage{xcolor}
\usepackage[export]{adjustbox}
\usepackage{amsmath}
\usepackage{amssymb}
\usepackage{chemformula,array}

\title{Geometrically-frustrated interactions drive structural\\ complexity in amorphous calcium carbonate}

\author{Thomas C. Nicholas$^1$, Adam E. Stones$^2$, Adam Patel$^1$, F. Marc Michel$^3$, Richard J. Reeder$^4$,\\Dirk G. A. L. Aarts$^2$, Volker L. Deringer$^{1,\ast}$ \& Andrew L. Goodwin$^{1,\ast}$}

\begin{document}

\maketitle

\begin{affiliations}
	\item Inorganic Chemistry Laboratory, Department of Chemistry, University of Oxford, Oxford, UK
	\item Physical and Theoretical Chemistry Laboratory, Department of Chemistry, University of Oxford,\\ Oxford, UK
	\item Department of Geosciences, Virginia Tech, Blacksburg VA, USA
	\item Department of Geosciences, Stony Brook University, Stony Brook NY, USA
\end{affiliations}

\begin{abstract}

Amorphous calcium carbonate (ACC) is an important precursor for biomineralisation in marine organisms. Among the key outstanding problems regarding ACC are how best to understand its structure and how to rationalise its metastability as an amorphous phase. Here, we report high-quality atomistic models of ACC generated by using state-of-the-art interatomic potentials to help guide fits to X-ray total scattering data. Exploiting a recently-developed inversion approach, we extract from these models the effective Ca$\boldsymbol\cdots$Ca interaction potential governing ACC formation. This potential contains minima at two competing distances, corresponding to the two different ways in which carbonate ions bridge Ca$^{2+}$-ion pairs. We reveal an unexpected mapping to the Lennard-Jones--Gauss (LJG) model normally studied in the context of computational soft-matter, with the empirical LJG parameters for ACC taking values known to promote structural complexity. In this way we show that both the complex structure of ACC and its resilience to crystallisation are actually encoded in the geometrically-frustrated effective interactions between Ca$^{\boldsymbol 2+}$ ions.

\end{abstract}

\section*{Introduction}

Calcium carbonate is relatively unusual amongst simple inorganic salts in that it precipitates from aqueous solution in a metastable hydrated, amorphous form.\cite{Gebauer_2008} ACC---with a nominal composition of CaCO$_3\cdot x$H$_2$O ($x\simeq 1$)---can be stabilised for weeks by incorporating dopants such as Mg$^{2+}$ or PO$_4^{3-}$,\cite{Aizenberg_2002,Raz_2003} or alternatively directed to crystallise into a number of different polymorphs by varying pH or temperature.\cite{Aizenberg_2004} Nature exploits this complex phase behaviour in a variety of biomineralisation processes to control the development of shells and other skeletal structures.\cite{Addadi_2003,Weiner_2005} Not only does the amorphous nature of biogenic ACC allow transformation to different crystalline CaCO$_3$ polymorphs, but it helps organisms fashion larger-scale hierarchical morphologies that are important in biomineral architectures.\cite{Beniash_1997} In seeking to develop bio-inspired crystal-engineering approaches for synthetic control over phase and morphology selection, there is an obvious need to understand why such a chemically simple system can exhibit such complex phase behaviour.

One domain in which a similar kind of phase complexity has been studied deeply from a theoretical perspective is that of soft-matter systems governed by multi-well pair potentials. Whereas isotropic particles that interact via a simple single-well potential (\emph{e.g.}\ Lennard-Jones, LJ) self-assemble into structurally simple crystalline phases (\emph{e.g.}\ face-centred cubic), the inclusion of one or more additional energy minima to the interaction potential can drive remarkable complexity if the distances at which these minima occur lead to geometric frustration.\cite{Rechtsman_2006,Engel_2008,Jain_2014,Dshemuchadse_2021} An elegant example in two dimensions is that of quasicrystal self-assembly for specific parameters of the double-well Lennard-Jones--Gauss (LJG) potential.\cite{Rechtsman_2006,Engel_2007} In three dimensions the same interaction model can be tuned to stabilise complex crystals with enormous unit cells,\cite{Dshemuchadse_2021} and some combinations of the LJ and Gauss well positions even appear to frustrate crystallisation altogether.\cite{Engel_2008}

That competing length-scales might be relevant to ACC is a point raised by the observation of two preferred Ca$\cdots$Ca distances dominating medium-range order in synthetic ACC.\cite{Michel_2008,Goodwin_2010} These two distances are directly evident in the experimental X-ray pair distribution function (PDF) of ACC and are attributed to different bridging modes of the carbonate ion, which can connect a pair of Ca$^{2+}$ ions either directly through one of the oxygen atoms (Ca--O--Ca pathway) or by inserting the carbonate ion fully between the cations (Ca--O--C--O--Ca pathway).

Here we explore the possibility that the structure of ACC is governed by effective interactions between Ca$^{2+}$ ions that also reflect these two length-scales, and that its well-known complexity emerges because the distances involved are in competition with one another. Our approach begins by obtaining a high quality measure of the Ca-pair correlation function in ACC. We do this by applying a hybrid Reverse Monte Carlo (HRMC) approach\cite{Opletal_2002} to generate the first structural model of ACC that is simultaneously consistent with experiment and stable with respect to state-of-the-art potentials. The Ca-pair correlation function that emerges is then inverted using a recently-developed algorithm\cite{Stones_2019} to reveal the effective, carbonate/water-mediated Ca$\cdots$Ca interaction potential. We find that this potential is closely related to the LJG formalism, with empirical parameters that are known to frustrate crystallisation. Monte Carlo (MC) simulations driven by our LJG model yield coarse-grained representations of ACC that capture key aspects of our fully-atomistic HRMC models. In this way, we explain the structural complexity of ACC in terms of geometric frustration of two competing energetically-favoured Ca$\cdots$Ca separations. Mapping the problem of ACC structure onto the phase behaviour of multi-well potentials is important not only because it establishes the first experimental system for which these potentials are relevant, but because it suggests how structural complexity in inorganic phases might be controllably targeted through suitable tuning of effective interactions.

\section*{Results}

\noindent{\bf Structure of ACC.} Our HRMC refinements made use of atomistic configurations containing 12\,960 atoms (1620 CaCO$_3\,\cdot\,$H$_2$O formula units), with simulation cell sizes of $\sim$5\,nm. During refinement, atomic moves were proposed and then rejected or accepted according to a Metropolis Monte Carlo criterion, with a cost function that measured the quality of fit to X-ray total scattering data\cite{Pusztai_1988} and also the energetic stability evaluated using the state-of-the-art interatomic potential of Ref.~\citenum{Raiteri_2010}. We favoured an HRMC approach over empirical potential structure refinement (EPSR) because calcium carbonate potentials are notoriously delicately balanced,\cite{Raiteri_2010} such that EPSR-derived potentials are unlikely to be physically meaningful for this system. To allow for direct comparison with previous studies, we used a $Q$-weighted variant of the experimental X-ray total scattering function (\emph{i.e.}\ $QF_{\rm X}(Q)$) of synthetic ACC reported in Ref.~\citenum{Michel_2008}, which has been shown to be indistinguishable from that obtained for biogenic ACC samples.\cite{Reeder_2013} For consistency, our HRMC configuration size and refinement constraints were also identical to those used in the reverse Monte Carlo (RMC) investigation of synthetic ACC in Ref.~\citenum{Goodwin_2010}. The absence of key energetic considerations in that RMC study allowed the development of physically unreasonable charge separation to give a model of cationic Ca$^{2+}$-rich domains separated by channels of carbonate anions and water molecules. Our expectation was that the explicit consideration of electrostatics in our HRMC implementation would guide refinement towards a solution that was equally consistent with experiment but also energetically sensible.

\begin{figure}
	\begin{center} 
		\includegraphics{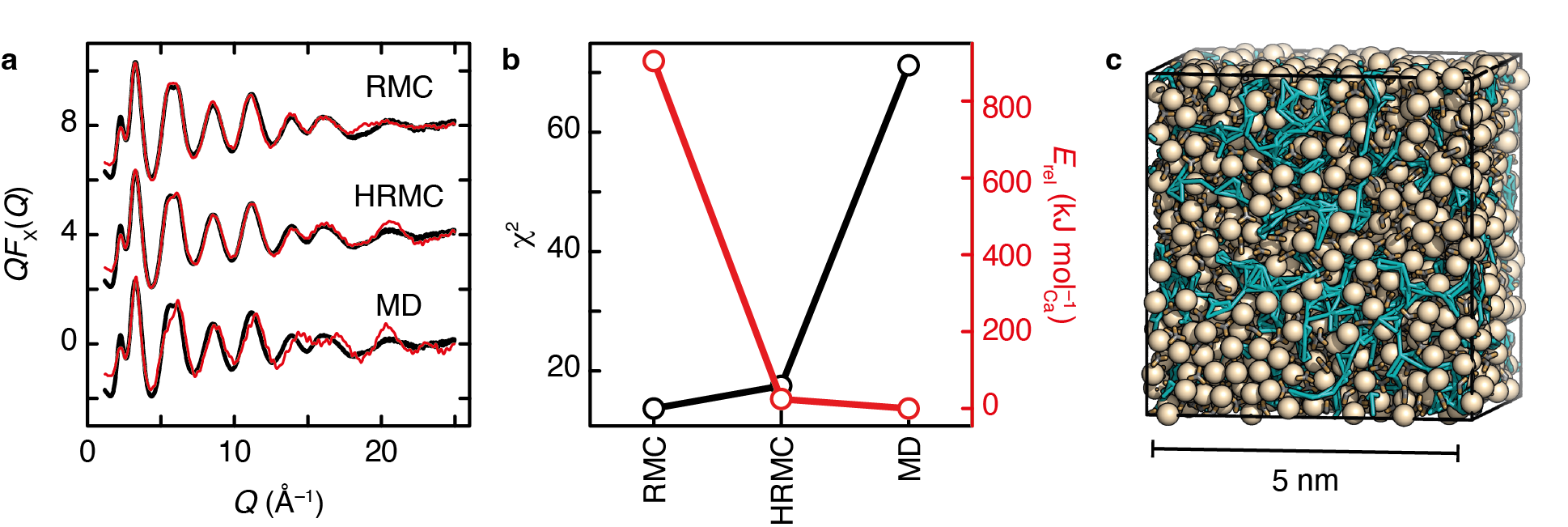}
	\end{center}
	\caption{\label{fig1}\footnotesize{\bf Hybrid reverse Monte Carlo yields a balanced model of ACC structure.} {\bf a} The experimental $Q$-weighted X-ray total scattering function $QF_{\rm X}(Q)$ is well fitted by both RMC (top) and HRMC (middle) refinements, but is meaningfully different to that calculated using the interatomic potential of Ref.~\citenum{Raiteri_2010} (bottom). Experimental data are shown as black lines, and fits or calculations as red lines. {\bf b} Quality criteria for the different structural models. HRMC simultaneously optimises both goodness-of-fit to experimental data $\chi^2$ and cohesive energy ($E_{\rm rel}$), and therefore it finds a structure solution that is essentially as consistent with experiment as that obtained using RMC, whilst also as energetically sensible as that obtained using potentials alone (MD). {\bf c} Representation of a converged structure of ACC obtained using HRMC. Water molecules connect to form filamentary strands (aquamarine strings) that separate calcium-carbonate-rich domains (Ca atoms as large beige spheres, carbonate ions shown as stick representation).} 
\end{figure}

HRMC does indeed find a suitable compromise between experiment and theory. We show in Fig.~\ref{fig1}(a) the $QF_{\rm X}(Q)$ function calculated from a representative configuration, and compare it to the equivalent functions predicted using pure RMC refinement, on the one hand, and unconstrained molecular-dynamics (MD) simulations with the potentials of Ref.~\citenum{Raiteri_2010}, on the other hand. Both RMC and HRMC give very similar high-quality fits to experiment---unsurprising, of course, since they have been refined against these data---whilst the MD simulation misses some aspects of the $QF_{\rm X}(Q)$ function. By contrast, the HRMC and MD models give similar cohesive energies, whereas the RMC model is less stable than both by more than 800\,kJ\,mol$^{-1}$ per formula unit. These results are summarised in Fig.~\ref{fig1}(b), which captures the motivation for our use of HRMC as a suitable balancing act: it has, for the first time, allowed access to an atomistic representation of ACC structure that is consistent with experiment and gives sensible energies using established hydrated calcium carbonate potentials. We checked also whether the HRMC model could reproduce the neutron total scattering data of Refs.~\citenum{Cobourne_2014,Wang_2017,Jensen_2018} (it can), and whether it is stable in molecular dynamics (MD) simulations driven by the potentials of Ref.~\citenum{Raiteri_2010} (it is); both points are discussed in greater detail in the Supplementary Material.

The HRMC model itself is illustrated in Fig.~\ref{fig1}(c). We observe that water is not homogeneously distributed throughout the configuration, but rather that the ACC structure consists of CaCO$_3$-rich regions separated by a filamentary network of water: we term this a `blue cheese' model. Qualitatively similar descriptions were obtained in MD simulations,\cite{Demichelis_2011} in EPSR refinements of combined neutron/X-ray total scattering data,\cite{Jensen_2018} and also inferred from solid-state NMR measurements.\cite{Ihli_2014} In all cases, the interpretation is that nearly all H$_2$O molecules are bound to Ca$^{2+}$ (as implied by $^{1}$H NMR; Ref.~\citenum{Nebel_2008}), but neighbouring water molecules are sufficiently close to form a network that percolates the ACC structure. As anticipated, the charge separation that developed in the RMC model of Ref.~\citenum{Goodwin_2010} has now vanished: the water-rich filaments we observe do not contain any free carbonate and are significantly narrower than the nanopore channels reported previously. The fact that RMC and HRMC models give similar $QF_{\rm X}(Q)$ fits despite their very different descriptions of the structure reflects the difficulty of discriminating between C and O atoms using X-ray scattering methods alone.\cite{Goodwin_2019}

\begin{figure}
	\begin{center} 
		\includegraphics{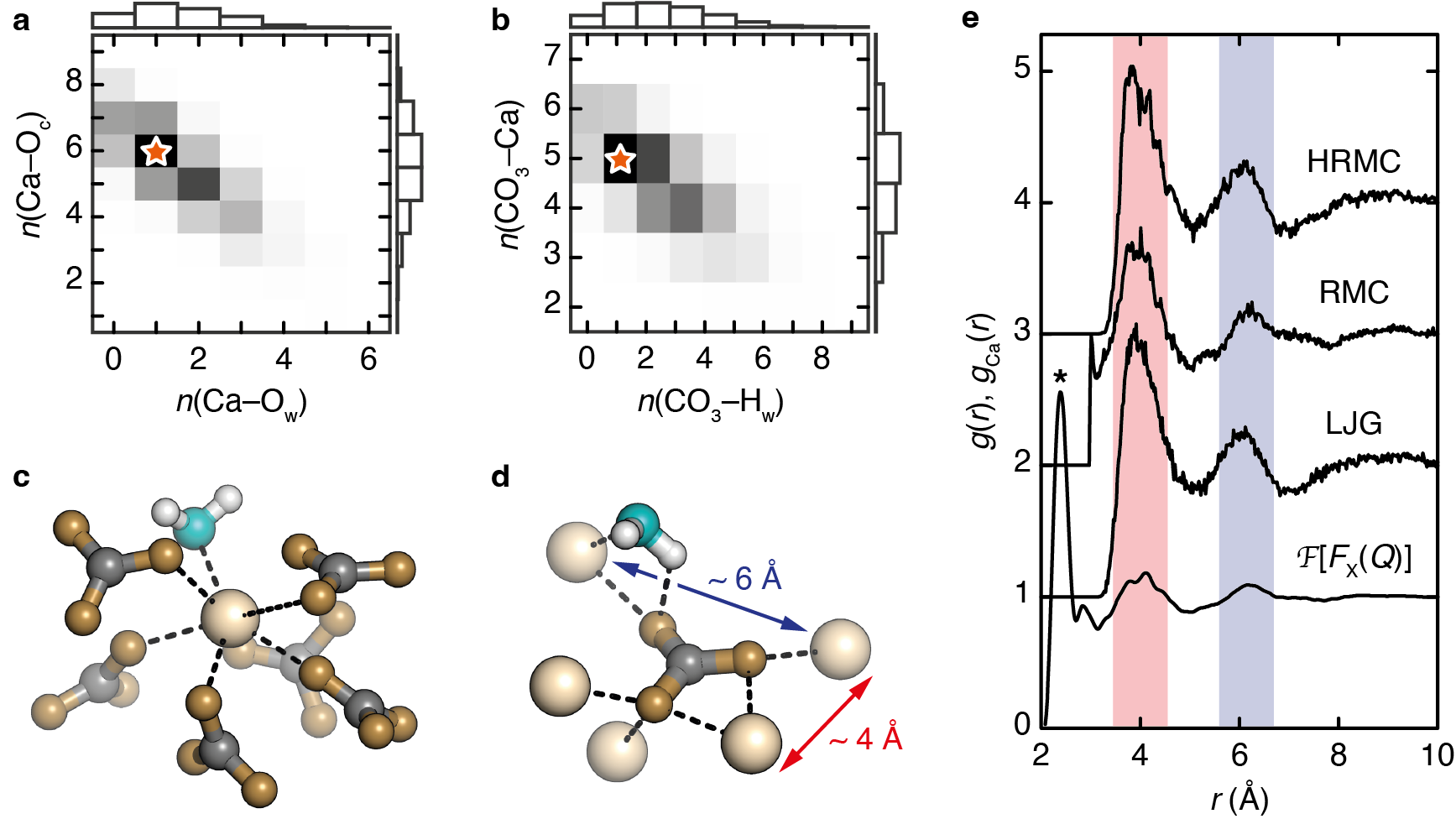}
	\end{center}
	\caption{\label{fig2}\footnotesize{\bf Coordination environments and Ca-pair distributions in ACC.} {\bf a} Histogram of Ca$^{2+}$ coordination environments, decomposed into contributions from carbonate and water oxygen donors (O$_{\rm c}$ and O$_{\rm w}$, respectively). {\bf b} Histogram of CO$_3^{2-}$ coordination environments, now decomposed into contributions from Ca$^{2+}$ and water hydrogen donors (O$_{\rm w}$). {\bf c} Representative Ca$^{2+}$ coordination sphere for the modal coordination environment marked by a star in {\bf a}. {\bf d} Representative CO$_3^{2-}$ coordination sphere for the modal coordination environment marked by a star in {\bf b}. Note that pairs of Ca$^{2+}$ ions within the same CO$_3^{2-}$ coordination sphere either share a common oxygen donor (\emph{e.g.}\ red arrow) or are connected by Ca--O--C--O--Ca pathways (\emph{e.g.}\ blue arrow). Ca, C, O, and H atoms shown in beige, grey, aquamarine, and white, respectively. {\bf e} Ca-pair correlation functions $g_{\rm Ca}(r)$ extracted from HRMC, RMC, and LJG configurations, compared against the normalised Fourier transform of the experimental X-ray total scattering function (which includes contributions from all atom-pairs, \emph{e.g.}\ the Ca--O peak at 2.4\,\AA\ marked with an asterisk). The two principal peaks common to all functions, indicated by red and blue shading, can be assigned to the two types of Ca$^{2+}$-ion pairs illustrated in {\bf d}.} 
\end{figure}

Locally, our HRMC model shows coordination environments that are consistent with the consensus of recent computational and experimental studies. For completeness, we show in Fig.~\ref{fig2}(a) and (b) the distribution of calcium and carbonate coordination numbers. The average Ca$^{2+}$ coordination number of 7.0---defined for a cut-off distance of 2.8\,\AA---is similar to that reported in Refs.~\citenum{Singer_2012,Jensen_2018,Clark_2022}. Whereas most Ca$^{2+}$ bind either five or six distinct carbonate anions, the carbonates tend to bind one fewer Ca$^{2+}$ ion each, reflecting a binding-mode distribution of $\sim$80:20 monodentate:bidentate. Representative modal coordination geometries for Ca$^{2+}$ and CO$_3^{2-}$ are shown in Fig.~\ref{fig2}(c) and (d). As anticipated,\cite{Goodwin_2010} we find that carbonate anions bridge calcium-ion pairs in two ways: either a pair of Ca$^{2+}$ ions share a common carbonate oxygen neighbour, or they bind distinct oxygens and so are connected formally by a Ca--O--C--O--Ca pathway. A full analysis of coordination environments, including bond-length and bond-angle distributions, is provided in the Supplementary Material.

\noindent{\bf Coarse graining.} In seeking to improve our understanding of the structure of ACC, we focussed on the Ca-pair correlation function $g_{\rm{Ca}}(r)$---after all, this is the contribution to the pair distribution function that exhibits the strongest persistent well-defined oscillations. In Fig.~\ref{fig2}(e), we show this function as extracted from our newly-generated HRMC configurations compared against that reported in the RMC study of Ref.~\citenum{Goodwin_2010}. We also include the Fourier transform of the experimental $F_{\rm X}(Q)$ function; this transform is an approximate (total) pair distribution function that emphasises contributions from Ca--Ca pairs as a consequence of the larger X-ray scattering cross-section for Ca relative to C, O, and H. While all three real-space correlation functions show maxima at $r\simeq4$ and $6$\,\AA, it is the HRMC result that resolves these features most clearly. By counting Ca--Ca pairs separated by Ca--O--Ca and Ca--O--C--O--Ca pathways in our HRMC configuration, we confirmed that the two maxima in $g_{\rm{Ca}}(r)$ occur at the preferred Ca$\cdots$Ca distances associated with these two different carbonate bridging motifs (see Supplementary Material).

\begin{figure}
	\begin{center} 
		\includegraphics{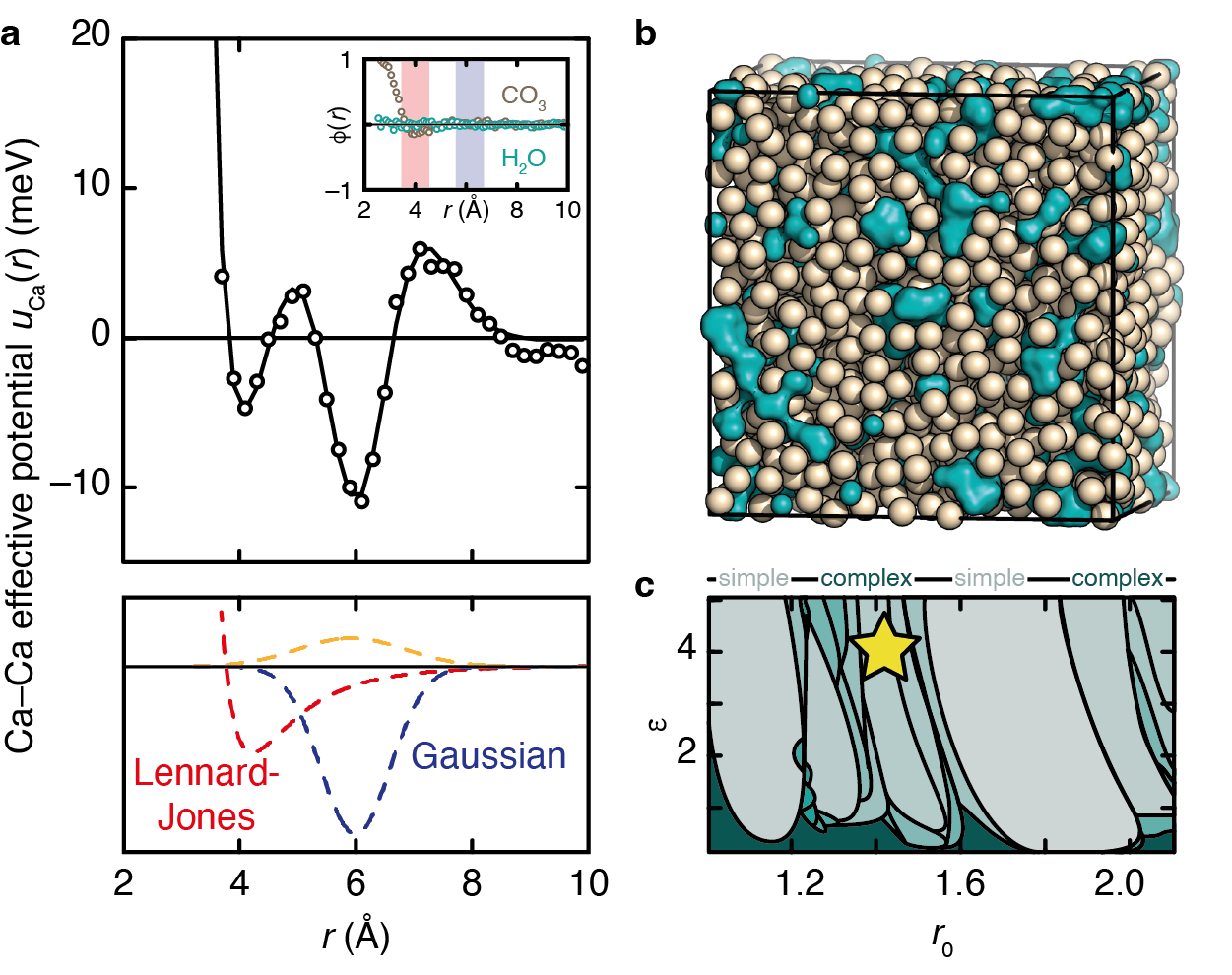}
	\end{center}
	\caption{\label{fig3}\footnotesize{\bf Effective Ca$\cdots$Ca interactions in ACC.} {\bf a} The effective Ca$\cdots$Ca interatomic potential extracted from our HRMC configurations\cite{Stones_2019} (open circles) and least-squares fit using a modified LJG model (line) as described in the text. The inset shows the orientational correlation functions (Eq.~\eqref{phi}), which vanish for distances relevant to the Ca$\cdots$Ca separations. {\bf b} Representative MC configuration of Ca atoms (beige spheres) generated by the LJG potential, parameterised by the fit shown in {\bf a}. The Ca atoms are not uniformly distributed, but cluster to leave Ca-poor voids, shown as aquamarine surfaces. Note the qualitative similarity to the heterogeneous structure of ACC represented in Fig.~\ref{fig1}(c). {\bf c} Approximate location of ACC effective LJG parameters in the LJG phase space as reported in Ref.~\citenum{Dshemuchadse_2021}.} 
\end{figure}

Access to a smoothly-varying measure of $g_{\rm{Ca}}(r)$, together with the HRMC configurations from which it is calculated, allowed us to exploit a recently-developed approach for the direct measurement of effective pair potentials from particle-coordinate data.\cite{Stones_2019} The method works by equating the pair distribution functions measured directly, on the one hand, and calculated using a test-particle insertion approach,\cite{Henderson_1983,Stones_2018} on the other hand.\cite{Widom_1963}  Applying this methodology to the atomic coordinates in our HRMC-refined model, we extracted the effective Ca$^{2+}$-ion pair potential $u_{\rm Ca}(r)$ shown in Fig.~\ref{fig3}(a). This function has two distinct potential wells, the minima of which are centred near distances for which $g_{\rm{Ca}}(r)$ has its maxima. The physical meaning of $u_{\rm Ca}(r)$ is that it captures the effective two-body interactions between Ca--Ca pairs, mediated by carbonate and water, as required to account for the observed $g_{\rm Ca}(r)$. That this interaction energy is minimised when Ca--Ca pairs are separated by distances corresponding to either Ca--O--Ca or Ca--O--C--O--Ca pathways makes intuitive sense (of course); the longer (6\,\AA) separation appears energetically more favourable likely because it minimises the electrostatic repulsion between calcium ions bound to a common carbonate. 

One assumption in using an isotropic effective pair potential to capture interactions mediated by anisotropic molecular species, such as H$_2$O and CO$_3^{2-}$, is that any local anisotropy is sufficiently short-ranged. We checked this point by calculating from our original (all-atom) HRMC configurations the orientational correlation functions
\begin{equation}
\phi(r)=\langle P_2(\mathbf S\cdot\mathbf S)\rangle_r,\label{phi}
\end{equation}
where $P_2(x)$ is the second-order Legendre polynomial, and the $\mathbf S$ are vectors parallel to a suitable local axis (\emph{e.g.}\ $C_3,C_2$) of molecules separated by distance $r$.\cite{Cinacchi_2019}  We find that $\phi(r)$ is essentially featureless---for both carbonate and water---for all but the very shortest intermolecular approaches, which occur for distances smaller than the nearest-neighbour Ca$\cdots$Ca separation [inset to Fig.~\ref{fig3}(a)]. Hence the anisotropy of individual molecules is relevant only at a length-scale smaller than that that over which the effective interactions between Ca$^{2+}$-ion pairs operate.

The double-well form of $u_{\rm Ca}(r)$ is important for a number of reasons. First, it explains why simple hard-sphere or LJ potentials fail to reproduce the experimental $g_{\rm Ca}(r)$: if tuned to capture the 4\,\AA\ nearest-neighbour correlation, such potentials always predict a second maximum just below 8\,\AA\ that is not observed in experiment. Second, it allows us to rationalise the inability of earlier EPSR studies to determine the correct Ca distribution function; for example, the study of Ref.~\citenum{Cobourne_2014} included a LJ parameterisation of Ca$\cdots$Ca interactions that then resulted in a $g_{\rm{Ca}}(r)$ function with maxima at positions clearly inconsistent with the inverse Fourier transform of the experimental $F^{\rm X}(Q)$. And, third, multi-well potentials are well known to support very complex phase behaviour,\cite{Engel_2007,Engel_2008,Elenius_2009,Mihalkovic_2012,Zhou_2019,Dshemuchadse_2021} suggesting a qualitative explanation of the complexity of ACC.

\noindent{\bf LJG parameterisation.} To test this hypothesis of a link between effective potential and structural complexity, we sought to map $u_{\rm Ca}(r)$ onto a suitable multi-well potential for which the corresponding theory is already well established. We found that a double-well LJG interaction\cite{Rechtsman_2006} described the observed functional form surprisingly well, so long as we included an additional broad, repulsive, Gaussian term that helps capture the local maxima between and beyond the two minima of the LJG function. The behaviour of the LJG potential is characterised by three parameters---$\epsilon,r_0,\sigma$---which describe, respectively, the depth, position, and width of the Gaussian well relative to the LJ component.\cite{Engel_2008,Dshemuchadse_2021} A least-squares fit to $u_{\rm Ca}(r)$ gave the values $\epsilon=4.1$, $r_0=1.4$ and $\sigma=0.14$ [Fig.~\ref{fig3}(a)]; note that $r_0$ is particularly well defined because it is closely related to the ratio of the Ca$\cdots$Ca separations resulting from the two different carbonate bridging motifs ($\simeq6$\,\AA\ / 4\,\AA). We will return to the importance of these empirical parameters in due course.
We note in passing related work on the description of coarse-grained molecular systems with double-well potentials, where one minimum is explicitly assigned to enthalpic and one to entropic terms,\cite{Pretti_2021} and on the `learning' of effective pair potentials from simulation data by back-propagation.\cite{Wang_2023}

As a simple check that our combination of $g(r)$ inversion and potential parameter fitting does indeed result in a meaningful effective pair-potential, we carried out direct MC simulations driven by our parameterised LJG potential. These simulations were performed using the experimentally-determined Ca particle density in ACC. The corresponding pair correlation function matches closely our HRMC $g_{\rm Ca}(r)$, as expected [Fig.~\ref{fig2}(e)], and the resulting structure resembles the Ca distribution in the fully atomistic model [Fig.~\ref{fig3}(b); \emph{cf}.~Fig.~\ref{fig1}(c)]. But the similarity between coarse-grained LJG-driven and HRMC Ca distributions turns out to extend beyond pair correlations, and we quantify this more extensive similarity in various forms (\emph{e.g.}\ ring statistics, Voronoi volume distributions, and higher-order correlation functions) in the Supplementary Material. A particular point of interest is that the MC configurations contain Ca-rich/-poor regions that are qualitatively similar to those observed in the original atomistic model [Fig.~\ref{fig3}(b)]---the clear implication being that, at the density of ACC, an inhomogeneous Ca distribution is encoded within the effective Ca$\cdots$Ca interaction itself.

\section*{Discussion}

From the perspective of soft-matter theory, much of the interest in the LJG potential lies in the complexity of phase behaviour which it supports.\cite{Engel_2008,Engel_2007,Suematsu_2014,Dshemuchadse_2021} This complexity arises because of competition between the structure-directing effects of LJ and Gaussian components, which operate at different length-scales. When the positions of the LJ and Gaussian minima are related by a factor $1.2\leq r_0\leq1.6$, there is particularly strong geometric frustration that results in a general resistance to crystallisation\cite{Engel_2008} and the emergence of many competing low-symmetry ground-state structures.\cite{Dshemuchadse_2021} This behaviour extends across a wide variety of relative well depths $\epsilon\geq1$ for the corresponding relative Gaussian width $\sigma=0.14$.\cite{Dshemuchadse_2021} Hence the empirical parameters we have determined to be relevant to effective Ca$\cdots$Ca interactions in ACC---\emph{viz.}\ $r_0=1.4, \epsilon=4.1,$ and $\sigma=0.14$---locate the system in one of the most complicated parts of the LJG phase diagram [Fig.~\ref{fig3}(c)].\cite{Dshemuchadse_2021} In developing this mapping onto the LJG model, we assume that the additional broad Gaussian term we have included in our parameterisation does not strongly influence phase behaviour (note that the value of $r_0$ does not vary significantly if it is omitted from our fit). Likewise, we cannot be certain of the effects of finite temperature and fixed density on LJG phase behaviour, since these have not yet been studied from theory. It is nevertheless a general phenomenon in frustrated systems that regions of strong geometric frustration are characterised by the existence of many competing ground states with suppressed ordering temperatures, above which the system is disordered but far from random.\cite{Ziman_1979,Moessner_2006}

In the context of ACC, the implications are twofold. First, the different favoured Ca$\cdots$Ca separations for carbonate-bridged Ca pairs---\emph{viz.}\ 4 and 6\,\AA---are configurationally difficult to satisfy in a three-dimensionally periodic (\emph{i.e.}, crystalline) structure with the density of ACC. This rationalises, in general terms, why an amorphous form of CaCO$_3$ is so relatively (meta)stable. And, second, the sensitivity of the LJG potential around $r_0=1.4$ suggests---again in general terms---a qualitative explanation of why ACC might be directed to crystallise into a number of different polymorphs. Of course, as the water content of ACC varies during the ageing of the material, both the effective Ca$\cdots$Ca interactions and the bulk density must change, which makes extrapolation from the LJG effective potential of ACC difficult. Increased material density also strengthens the orientational correlations in molecular components, and so one expects the isotropic effective pair potential description to break down as crystallisation is approached. Nevertheless, at the density of ACC and in the absence of orientational order, the relative stability of a disordered state is now more easily rationalised. Likewise, the incorporation of Mg$^{2+}$ or PO$_4^{3-}$ ions into ACC will introduce statistical variations into the effective potential governing cation arrangements, stabilising further the amorphous state---much as magnetic exchange disorder can stabilise spin-glasses.\cite{Saunders_2007}

Whereas the (hitherto unsuccessful) search for `real-world' materials governed by isotropic multi-well potentials has focussed on the atomic (alloys) and mesoscopic (colloids) length-scales,\cite{Engel_2007,Dshemuchadse_2021} our study of ACC suggests that the intermediate domain of nanoscale materials may in fact be a more fertile source of relevant examples. By their very nature, molecular ions tend to exhibit a range of coordination modes, which must give rise to multiple specific preferred distances between the counter-ions they coordinate. Whenever these distances are geometrically frustrated, as in ACC, one expects that the corresponding effective potential must contain multiple wells so as to stabilise both separations at once. Through suitable choice of chemical components, one might hope to navigate the complicated phase space associated with such unconventional potentials. Doing so will provide an important test of theory, on the one hand, and also allow the targeted design of complex structures through self-assembly, on the other hand.

Our coarse-grained approach to understanding ACC structure may be applicable to other poorly ordered inorganic solids of particular scientific importance. Amorphous calcium phosphate (the precursor to bone) and calcium--silicate--hydrates (\emph{i.e.}\ Portland cements) are obvious ever-contentious examples,\cite{Lowenstam_1985,Richardson_2014} where one might expect the different bridging motifs---now of the phosphate or (poly)silicate anions---to again moderate an effective interaction potential with multiple minima. Better understanding the structures of long-studied materials is certainly one fruitful avenue for future research, but our work suggests also a new pathway towards complex materials design. By varying the inorganic cation radius for a given molecular counter-ion, or by changing the degree of hydration, one might hope to control the ratio of the distances at which minima in the effective potential occur: what emerges is a strategy of `interaction engineering' to target a particular phase of interest, or to stabilise or destabilise an amorphous form of matter.

\section*{Acknowledgments}
T.C.N. was supported through an Engineering and Physical Sciences Research Council DTP award (grant number EP/T517811/1).
A.L.G. gratefully acknowledges useful discussions with M.~T.~Dove (QMUL) and financial support from the E.R.C. (Grant 788144). The authors would like to acknowledge the use of the University of Oxford Advanced Research Computing (ARC) facility in carrying out this work (http://dx.doi.org/10.5281/zenodo.2255).


\end{document}